\documentclass[preprint,showpacs,preprintnumbers,amsmath,amssymb]{revtex4}
\usepackage{dcolumn}
\usepackage{bm}
\usepackage{amssymb,amsmath, amsthm,epsfig}

\begin{document}
\title{Forward and inverse cascades in decaying two-dimensional electron magnetohydrodynamic turbulence}
\author{C. J. Wareing\footnote{E-mail: cjw@maths.leeds.ac.uk}, R. Hollerbach}
\affiliation{Department of Applied Mathematics,
University of Leeds, Leeds, LS2 9JT, UK}

\begin{abstract}
Electron magnetohydrodynamic (EMHD) turbulence in two dimensions is studied via 
high-resolution numerical simulations with a normal diffusivity. The resulting 
energy spectra asymptotically approach a $k^{-5/2}$ law with increasing $R_B$,
the ratio of the nonlinear to linear timescales in the governing equation. No 
evidence is found of a dissipative cutoff, consistent with non-local 
spectral energy transfer. Dissipative cutoffs found in previous studies
are explained as artificial effects of hyperdiffusivity. 
Relatively stationary structures are found to develop in time, rather than 
the variability found in ordinary or MHD turbulence. Further, EMHD turbulence
displays scale-dependent anisotropy with reduced energy transfer in the 
direction parallel to the uniform background field, consistent with 
previous studies. Finally,  
the governing equation is found to yield an inverse cascade, at least 
partially transferring magnetic energy from small to large scales.
\end{abstract}


Copyright 2009 American Institute of Physics. This article may be downloaded
for personal use only. Any other use requires prior permission of the author
and of the American Institute of Physics.\\

The following article appearing in Physics of Plasmas, {\bf 16}, 042307 (2009)
and may be found at (http://link.aip.org/link/?PHP/16/042307).\\
 
\maketitle

\vspace{0.1cm}

\section{Introduction}

Turbulence plays a crucial role in a wide variety of geophysical and
astrophysical fluid flows.  In this paper we present results on a
particular type of plasma turbulence in which the flow is entirely due
to the electrons, with the ions forming a static background.  The
equation governing the electrons' self-induced magnetic field is then
\begin{equation}
\frac{\partial {\bf B}}{\partial t} =
 - \nabla \times \left[ {\bf J} \times {\bf B} \right] 
+ R_B^{-1} \ \nabla^2 {\bf B},
\label{eq:A}
\end{equation}
where ${\bf J} = \nabla \times {\bf B}$, and $R_B={\sigma B_0}/{n e c}$,
with $\sigma$ the conductivity, $B_0$ a measure of the field strength,
$n$ the electron number density, $e$ the electron charge, and $c$ the
speed of light.  See for example \cite{goldreich92}, who derived this
equation in the context of magnetic fields in the crusts of neutron
stars.  More generally though, it is applicable in many weakly
collisional, strongly magnetic plasmas, so other applications could
include the Sun's corona or the Earth's magnetosphere.

Turbulence governed by (\ref{eq:A}) is known as electron MHD (EMHD), Hall MHD,
or whistler turbulence.  Based on its (at least superficial) similarity
to the vorticity equation governing ordinary, nonmagnetic turbulence,
\begin{equation}
\frac{\partial {\bf w}}{\partial t} =
\nabla \times \left[ {\bf u} \times {\bf w} \right]
 + Re^{-1} \ \nabla^2 {\bf w},
\label{eq:B}
\end{equation}
where now ${\bf w} = \nabla \times {\bf u}$, \cite{goldreich92}
argued that (\ref{eq:A}) would initiate a turbulent cascade to small
lengthscales, thereby accelerating neutron stars' magnetic field
decay beyond what ohmic decay acting on large lengthscales could
achieve.  They suggested in particular that the turbulent spectrum
would scale as $k^{-2}$, with a dissipative cutoff occurring at
$k\sim R_B$.

However, there are also
some quite fundamental differences between equations (\ref{eq:A}) and
(\ref{eq:B}).  In (\ref{eq:B}) the dissipative term contains more
derivatives than the nonlinear term, so on sufficiently short
lengthscales the dissipative term will always dominate.  In contrast,
in (\ref{eq:A}) the two terms both contain two derivatives, so it is
conceivable that the nonlinear term will always dominate, even on
arbitrarily short lengthscales.  As pointed out by \cite{hollerbach02},
one obtains a dissipative cutoff only if one assumes that the cascade
is local in Fourier space, coupling wavenumbers only to their immediate
neighbors.

Indeed, whether the coupling is local or not is another important
difference between (\ref{eq:A}) and (\ref{eq:B}).  In (\ref{eq:B}) it is at
least predominantly local; for example, very small scale structures see
the largest structures as an essentially uniform background flow that
simply advects them along, but without altering the nature of the small
scale turbulence.  In contrast, in (\ref{eq:A}) there is no such
translational invariance; adding even an exactly uniform background
field alters the dynamics of the small scale structures (as we will show
in detail below). Similarly, in classical MHD turbulence, adding a background
flow has no essential effect, but adding a background field does.

In this paper we present high-resolution numerical simulations of
(\ref{eq:A}) in a two-dimensional periodic box geometry, designed
specifically to address such questions as whether there is a dissipative
cutoff or not, and whether the coupling is local or not.  In contrast
to previous simulations \cite{biskamp96,biskamp99,dastgeer00,dastgeer03,cho04,shaikh05}, we
do not employ hyperdiffusivity, which would of course disrupt this
feature that the two terms in (\ref{eq:A}) have the same number of
derivatives, and hence introduce an artificial dissipative cutoff.
We also consider the question of whether (\ref{eq:A}) is capable of
yielding an inverse cascade, and find that magnetic energy can be at
least partially transferred from small to large scales.

\section{Equations}

For two-dimensional fields, we may decompose $\bf B$ as
\begin{equation}
{\bf B} = {\bf B}_p + {\bf B}_t
 = \nabla \times (a \hat{{\bf e}}_z) + b \hat{{\bf e}}_z.
\label{tp1}
\end{equation}
where $a$ and $b$ depend only on $x$, $y$, $t$, but not $z$. 
Equation (\ref{eq:A}) then yields
\begin{equation}
\frac{\partial a}{\partial t} =
 R_B^{-1} \nabla^2 a - (a_y b_x - a_x b_y),
\label{tp2}
\end{equation}
and
\begin{equation}
\frac{\partial b}{\partial t} =
 R_B^{-1} \nabla^2 b + (a_y \nabla^2 a_x - a_x \nabla^2 a_y),
\label{tp3}
\end{equation}
where subscripts indicate derivatives.

Continuing our comparison of equations (\ref{eq:A}) and (\ref{eq:B}), it
is instructive to note also that in (\ref{eq:B}) we would only have
${\bf u}={\bf u}_p=\nabla \times (\Psi \hat{{\bf e}}_z)$, yielding
\begin{equation}
\frac{\partial \nabla^2 \Psi}{\partial t} = Re^{-1} \nabla^4 \Psi
 - [\Psi_y \ ( \nabla^2 \Psi)_x - \Psi_x \ ( \nabla^2 \Psi)_y ].
\label{psi}
\end{equation}

Any additional ${\bf u}_t=v \hat{{\bf e}}_z$ would simply be advected
by $\Psi$ as a passive scalar, but without any influence back on
$\Psi$.  This difference between (\ref{tp2}) and (\ref{tp3}) on the one
hand, and (\ref{psi}) on the other merely reflects once again some of
the differences between (\ref{eq:A}) and (\ref{eq:B}), in this case the
lack of translational invariance in (\ref{eq:A}).

We solve (\ref{tp2}) and (\ref{tp3}) by expanding $a$ and $b$ in
Fourier series in $x$ and $y$, and using standard pseudospectral
techniques for the evaluation of the nonlinear terms, with dealiasing
according to the 2/3 rule.  The code employs the FFTW library
\cite{frigo05} to achieve massive parallelisation on a suitable
supercomputer. The time integration is done using a second order
Runge-Kutta method. We performed a variety of runs, typically employing 64
processors, with the highest extending to $k=682$ in Fourier space,
corresponding to $N=2048$ collocation points in real configuration space.  Because
of the two derivatives in the nonlinear term, the required timesteps
are unfortunately very small, roughly proportional to $1/(N^2)$.  Values
as small as $\sim3\times 10^{-8}$ were used, requiring $O(10^7)$
timesteps in total to reach $t=0.2$.

\subsection{Initial Conditions}

Since we are interested in freely decaying rather than forced
turbulence, we need to carefully consider the nature of our chosen
initial conditions.  We will present results for three different
sets of runs.

First, to study homogeneous forward cascades, we start off with
random $O(1)$ energies in all Fourier modes up to $k=
\sqrt{k_x^2+k_y^2}=5$, making sure that the poloidal $a$ and
toroidal $b$ components have comparable amounts of energy.  After
initialisation the overall amplitude of the field is rescaled to
ensure that the rms value of $|{\bf B}|=1$ at $t=0$.

Second, to study nonhomogeneous forward cascades, we start off with
the same initialisation as above, but now add a uniform field
$C\hat{\bf e}_x$, where $C=1$, 2 or 4.  Note though that such a
uniform field cannot be represented by an expansion of the form
(\ref{tp1}), at least not if $a$ and $b$ are to be periodic in $x$ and
$y$.  Instead, this field is simply added in to (\ref{tp1}) directly,
resulting in suitably modified equations (\ref{tp2}) and (\ref{tp3}).

Third, to explore the possibility of inverse cascades, we return to
the $C=0$ case without any large scale magnetic field and now inject 
energy into modes in the range $10\le k\le 20$.  The question then is 
how much of this initial energy moves to $k<10$, and how much moves 
to $k>20$.

Finally, for all three sets of results, each individual run was
repeated with a number of different random initial conditions, to
ensure that the results presented here are indeed representative.

\subsection{Ideal Invariants}

Equations (\ref{tp2}) and (\ref{tp3}) also have some useful associated
diagnostics, corresponding to quantities that are conserved in the
ideal, $R_B^{-1}\to0$, limit.  Specifically, we have equations for
the energy and the magnetic helicity,
\begin{equation}
\frac{d}{dt} { \frac{1}{2} \int {\bf B}^2\,{\rm d}V } 
= -R_B^{-1} \int {\bf J}^2\,{\rm d}V,
\label{q1}
\end{equation}
\begin{equation}
\frac{d}{dt}{\frac{1}{2}\int{\bf A\cdot B}\,{\rm d}V}
=-R_B^{-1}\int{\bf B\cdot J}\,{\rm d}V,
\label{q2}
\end{equation}
where $A$ is the vector potential, defined by ${\bf B=\nabla\times A}$.
Note though that in the presence of a uniform background field,
helicity is not even defined \cite{berger97}, let alone conserved.

These two equations are valid in both 2 and 3 dimensions.  In 2
dimensions only, we have the additional quantity of the mean squared
magnetic potential, known as anastrophy
\begin{equation}
\frac{d}{dt}{\frac{1}{2}\int a^2\,{\rm d}V}=-R_B^{-1}\int(a_x^2 + a_y^2)\,{\rm d}V,
\label{q3}
\end{equation}
which is in some ways perhaps analogous to enstrophy, which is also
defined only for (\ref{psi}) in 2D, but not for (\ref{eq:B}) in 3D.
However, anastrophy is \emph{not} the same as enstrophy, and there does not
appear to be any reason why conservation of anastrophy would
necessarily imply an inverse cascade in the way that conservation of
enstrophy forces inverse cascades to exist in 2D hydrodynamic
turbulence.

In addition to the physical insight that they yield into the nature
of the Hall nonlinearity, these various integrated quantities also
offer useful diagnostic checks of the code.  Reassuringly, we found
that all of them (except helicity in a uniform field of course) were
satisfied to within 0.1\% or better by all of our runs.

\section{Results}

\subsection{Large-scale initial conditions}

The energy spectrum, $E_k$, of 3D hydrodynamic turbulence is characterised
by $E_k \propto k^{-5/3}$, the familiar Kolmogorov law 
\cite{kolmogorov41}. In 2D, the conservation
of enstrophy forces an inverse cascade which leads to a much
steeper spectrum $E_k \propto k^{-\alpha}$ with $\alpha > 3$.
In the case of 2D MHD turbulence, the spectral energy transfer
rate is reduced which leads to a flatter energy spectrum
$E_k \propto k^{-3/2}$, the Iroshnikov-Kraichnan spectrum \cite{kraichnan80}.
Recent studies of EMHD turbulence have found, via methods which
all employ hyperdiffusivity, a 5/3 Kolmogorov spectrum
for small scales $k d_e > 1$, equivalent to $k > O(R_B)$, and 
a steeper 7/3 spectrum for longer wavelengths
\cite{biskamp96,biskamp99,dastgeer00,dastgeer03,cho04}.

\begin{figure}
\begin{center}
\includegraphics[angle=0,width=8cm]{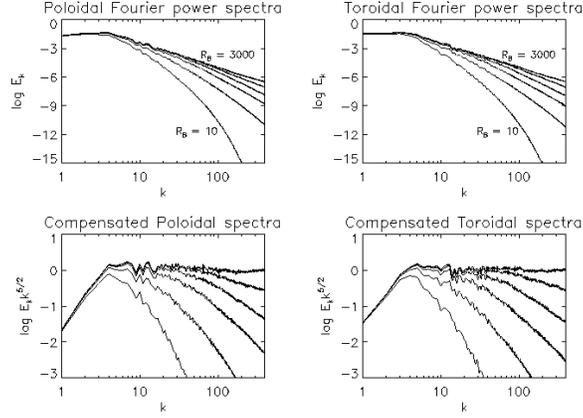}
\caption{Energy spectra of homogeneous 2D EMHD turbulence at $t=0.2$. 
Across the top row are shown spectra for both toroidal and
poloidal fields for $R_B=10$, 30, 100, 300, 1000 \& 3000. Across the 
bottom row are compensated energy spectra $k^{5/2} E_k$.}
\label{cascades}
\end{center}
\end{figure}

In the upper plot of Figure \ref{cascades}, we show the poloidal
and toroidal energy spectra of our solutions for $R_B = 10$, 30, 100, 300,
1000 \& 3000, evolved to a time $t = 0.2$. The energy spectra
have been stationary since approximately $t = 0.12$ and time 
averaging between 0.12 and 0.2 reveals an identical spectrum
and no further information. We interpret this to mean
our simulations are resolved and evolved to a suitable time
for inspection of the quasi-stationary cascade.

Both poloidal and toroidal energy spectra start out much the 
same at low $k$ and then lower $R_B$ spectra smoothly drop off 
with increasing $k$ whilst higher $R_B$ spectra maintain a linear
gradient in the log-log plot. Transfer of energy to higher $k$ is then more efficient at
higher $k$. The spectra are asymptotically approaching an energy 
spectrum $E_k \propto k^{-\nu}$, where $\nu = 2.5 \pm 0.1$. In 
the lower plot of Figure \ref{cascades}, we show compensated 
energy spectra to show this approach to $k^{-5/2}E_k = 1$ with
increasing $R_B$. Our value of $\nu$ is not compatible with $-2$ 
predicted by \cite{goldreich92} for 3D EMHD 
turbulence. Their prediction was calculated using a phenomenology
based on Kraichnan's arguments (the whistler effect) which has 
been shown by \cite{dastgeer00} to have little effect on the 
energy spectrum of 2D EMHD, rather it is thought to influence the subtle properties of the
cascade like anisotropy which we discuss further in the next section.
The spectral index we find is much more compatible with $\nu = 7/3$ 
found by \cite{biskamp99} via numerical simulation for the $k d_e > 1$ 
regime. 

None of the spectra show any sign of a dissipative cutoff. 
By definition, the dissipation scale should occur when the 
local value of $R_B$ is $O(1)$ in equation \ref{eq:A}. It is unclear
though when this occurs since the definition of $R_B$
does not involve length scales. If the coupling is 
purely local in wavenumber, then this definition does involve
length scales after all, since the $B_0$ that should be used
is the field at that wavenumber only, rather than the total field.
That is, according to the definition of \cite{hollerbach02} where this
argument was first developed, we have
\begin{equation}
R_B' = R_B (B'/B)
\end{equation}
where the primed quantities are the small-scale local values and
the unprimed the large-scale global. If we now suppose
a $k^{-5/2}$ energy spectrum, then $B'/B \sim k^{-5/4}$ and so
$R_B'$ is reduced to $O(1)$ when $k \sim R_B^{4/5}$. So, at $R_B = 1000$
we would see a dissipative cutoff at $k \sim 250$. We do not see 
a dissipative cutoff at this scale. We can reconcile this by realising
that this argument crucially 
depends on the coupling being local in Fourier space: if this 
does not hold then $R_B' = R_B$ and there is simply no definite 
dissipation scale, as we find here. 

We conclude therefore that the nonlinear term 
is able to dominate at all length scales and the coupling is 
non-local in Fourier space. This is not entirely unexpected, since 
both the terms in (\ref{eq:A}) contain two derivatives, but it
is in contrast to previous studies. It should be emphasized 
that other authors have been unable to properly address the question 
of a dissipative cutoff since hyperdiffusivity has 
masked the effect of the nonlinear term at high $k$. This may
also explain the discrepancy between the spectral index of 5/2 we find
and previous values of 7/3. 

\begin{figure}
\begin{center}
\includegraphics[angle=0,width=16cm]{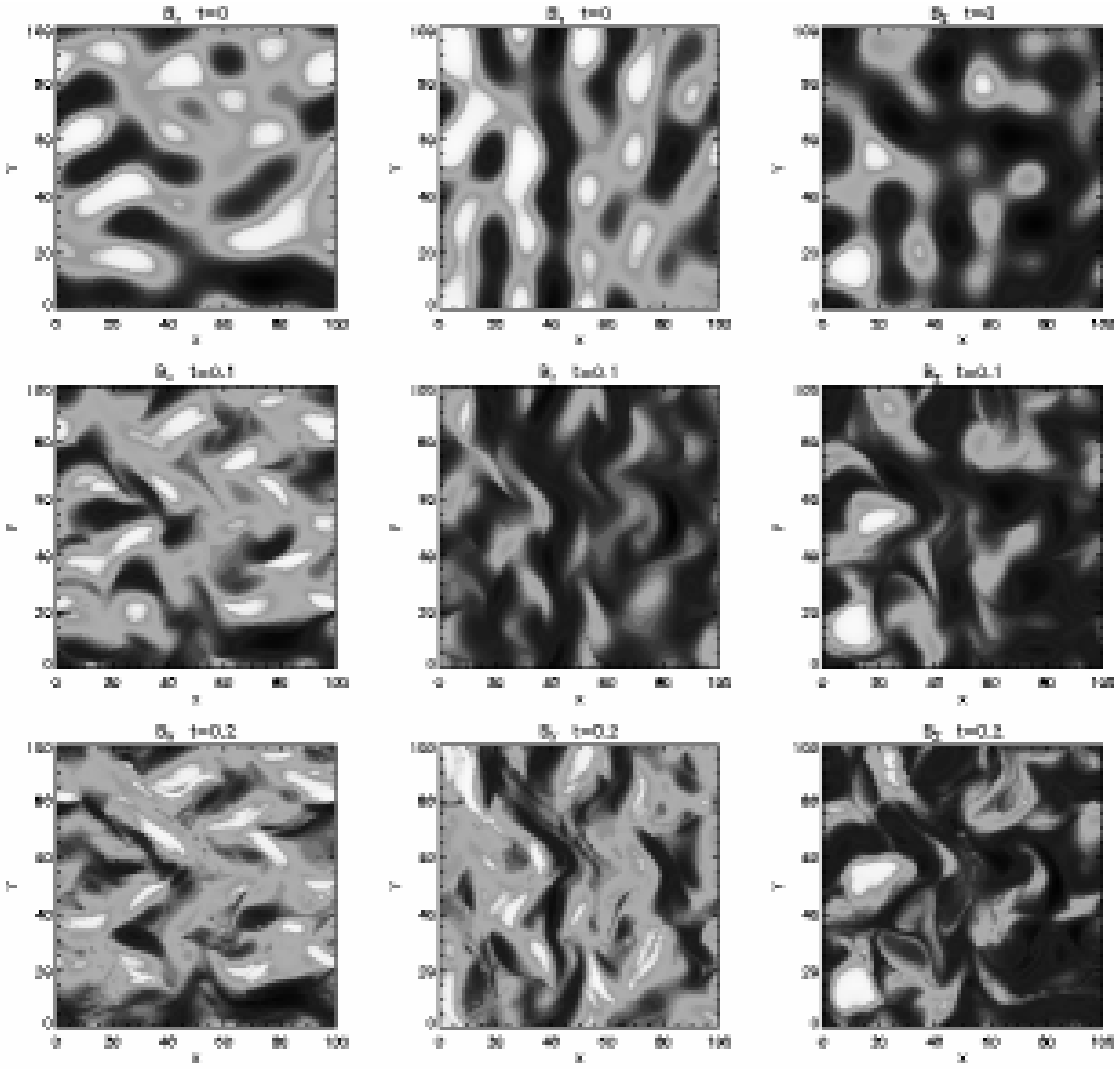}
\caption{Plots of the $R_B = 1000$ solution in real
configuration space at $t = 0$ (top row), $t=0.1$ 
(middle row) and $t = 0.2$ (bottom row).The fields
have been rescaled onto grids of $100 \times 100$
points.}
\label{truefield}
\end{center}
\end{figure}

In Fourier space then, EMHD turbulence bears a strong
resemblance to ordinary MHD turbulence. We would like to know
if this resemblance carries over into real configuration space. In
Figure \ref{truefield}, we show the three component fields at
three times; the initial fields at $t=0$ (top row), intermediate 
fields at $t=0.1$ (middle row) and fully developed turbulent 
fields at $t=0.2$ (bottom row). In classical and MHD turbulence, 
a fully developed turbulent field in real configuration space would bear no 
resemblance to the initial field, but here the fields are much 
more structured and fully developed turbulent fields strongly
resemble initial fields. This appears to be a unique characteristic of 
decaying EMHD turbulence.

\begin{figure}
\begin{center}
\includegraphics[angle=0,width=8cm]{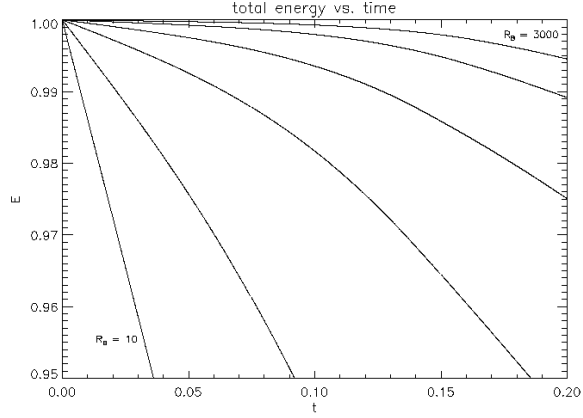}
\caption{A plot of energy against time for 
$R_B=10$, 30, 100, 300, 1000 \& 3000, increasing from
left to right.}
\label{energy}
\end{center}
\end{figure}

We would also like to address the energy decay of the field, with
particular respect to any dependency of the decay rate on the
value of $R_B$. Ref.\ \cite{biskamp96} in the first study of 2D 
EMHD reported that the energy dissipation rate is independent of the
value of the dissipation coefficient, represented by $R_B$ here.
In contrast to this, we find that the energy decay is much slower 
at higher $R_B$, as plotted in Figure \ref{energy}.

\subsection{Large-scale initial conditions in the presence 
of a background field}

EMHD turbulence, like classical and MHD turbulence, is 
isotropic when allowed to freely decay. In the presence of a 
background flow, classical turbulence remains isotropic since 
it is locally coupled in Fourier space. Small-scale structures 
are simply advected along by the large-scale flow, whether or 
not that has a uniform background contribution. Numerical 
simulations of MHD turbulence have found it to be strongly 
anisotropic in the presence of a background field 
\cite{shebalin83,oughton98}. This has been attributed to the
excitation of Alfv{\' e}n waves which preferentially propagate
parallel to the external magnetic field and hinder the cascade
process perpendicular to the external field.

In EMHD turbulence, recent numerical studies employing
hyperdiffusivity \cite{dastgeer00,dastgeer03} have revealed similar
strongly anisotropic behaviour. This can only be the result of
asymmetry in the nonlinear spectral transfer process relative
to the external magnetic field. In the context of local energy 
coupling in Fourier space, mediation by whistler waves has
been proposed as the only way this asymmetry could be achieved
\cite{dastgeer00} with the method detailed in ref.
\cite{galtier06}. The spectrum of 2D anisotropic EMHD
turbulence has also been shown to exhibit a linear relationship with
an external magnetic field \cite{dastgeer03}. 

\begin{figure}
\begin{center}
\includegraphics[angle=0,width=16cm]{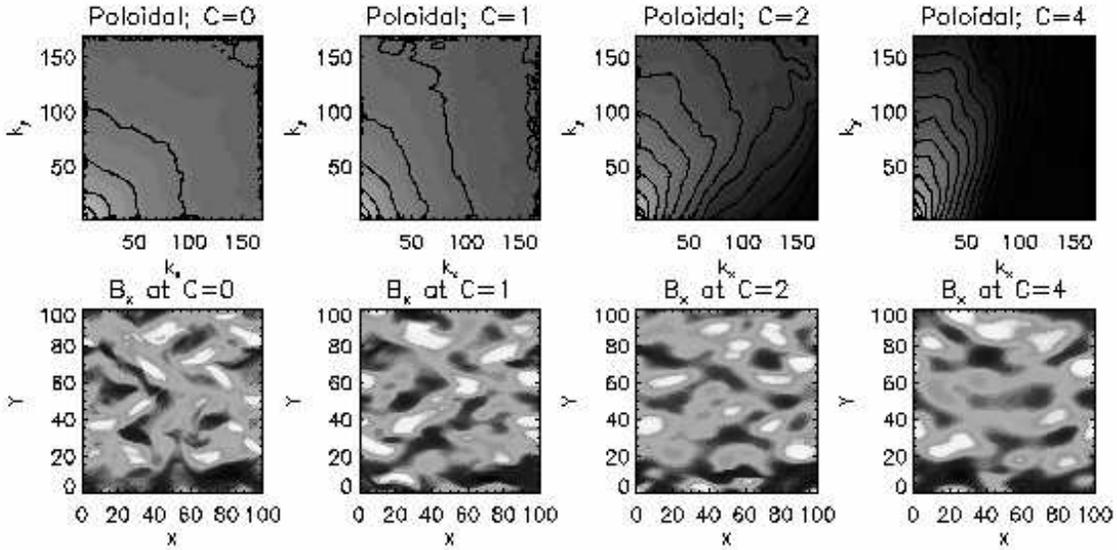}
\caption{Fourier power spectra and real space fields at $t=0.2$ 
for 2D EMHD turbulence in the presence of a background field. 
We show 2D Fourier power spectra of $a$, the poloidal field,
for $R_B = 100$ at $C=0$, 1, 2 \& 4 across the top row. Across the
bottom row we show the corresponding B$_x$ fields in real 
configuration space.}
\label{background1}
\end{center}
\end{figure}

In order to understand how hyperdiffusivity has affected previous
studies we have introduced a background field into the governing 
equations as discussed above and calculated solutions for $R_B = 100$ 
\& 300 at a spatial resolution of $512^2$ points in real configuration 
space. We present our results in Figure \ref{background1}. 
Across the top row, we show 2D energy spectra for $R_B = 100$ with 
$C=0$, 1, 2 \& 4. In the isotropic case with no background field, 
i.e. $C=0$, energy is evenly distributed between $x$ and $y$, as 
indicated by circular contours. In the case of $C=1$ we find energy
transfer to larger $k$ has been suppressed in the $x$ direction, 
parallel to the background field. 
EMHD turbulence has become anisotropic in the presence of a uniform 
background field with normal diffusivity. The effect becomes more 
pronounced for $C=2$ and $C=4$. The evolution of modes parallel and
perpendicular to the field is clearly different, as the spectral
cascade in the parallel wavenumbers is clearly suppressed. This
suppression has been attributed to excitation of whistler waves,
which act to weaken spectral transfer along the direction of
propagation \cite{dastgeer03}. Across the bottom row, we show the
corresponding $B_x$ fields. For $C=0$, the field is isotropic, but
as the value of $C$ is increased, structures are stretched in the $x$ direction
corresponding to increasingly inhibited energy transfer in $x$ but not in $y$,
perpendicular to the field direction.

This result is in agreement with \cite{cho04} who found scale-dependent 
anisotropy in numerical studies of 3D EMHD turbulence employing 
hyperdiffusivity. Our simulations also support the linear relationship
between 2D EMHD turbulence and strength of external magnetic field
found by \cite{dastgeer03}. At $R_B = 300$ we reassuringly find the 
field is more anisotropic.

It is worth noting here that since (\ref{eq:A}) is scale 
invariant - we can apply the equation over the whole of a system, or
just a small section, with $R_B$ unchanged - if you take a very small
box, then this box will see the large-scale field as a background field,
and therefore one would expect the smallest scales in the system,
for example a neutron star, to be anisotropic.

\subsection{Intermediate-scale initial conditions}

In classical turbulence, the exchange of energy and enstrophy 
$\Omega$ is coupled in Fourier space according to
\begin{equation}
\frac{\partial E}{\partial t} = - k^2 \frac{\partial \Omega}
{\partial t},
\end{equation}
hence energy injected at intermediate scales experiences a transfer 
to both higher and lower wavenumbers in order to satisfy this coupling 
and simultaneously conserve energy and enstrophy. This is the inverse 
cascade of energy to lower $k$ (larger scales) \cite{kraichnan67}. In MHD 
turbulence, energy and magnetic helicity are coupled in the same way
and an inverse cascade occurs in order to simultaneously conserve 
these two quadratic ideal invariants. To investigate whether an
inverse cascade occurs in decaying EMHD turbulence, we 
inject energy over the wavenumber range $10 \leq k \leq 20$ 
as detailed above and evolve the magnetic field. 

\begin{figure}
\begin{center}
\includegraphics[angle=0,width=8cm]{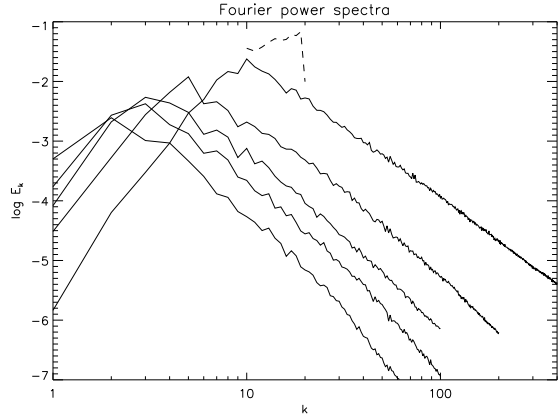}
\caption{Evolution of the Fourier power spectra for $R_B=1000$,
showing an inverse cascade of energy to $k<10$. Spectra 
are shown at $t=0.0$ (dashed line), 0.1, 1.0, 3.0, 6.0 
\& 15.0. Note the changes in resolution from $N=2048$ at 
$t=0.1$ to $N=1024$ at $t=1.0$ to $N=512$ at later times.}
\label{inverse}
\end{center}
\end{figure}

In Figure \ref{inverse} we show the solution for $R_B = 1000$ at 
various times. To fully resolve the solution in a reasonable amount 
of computational time we have chosen to evolve the field to $t=0.2$ 
at a resolution of $2048^2$ real space points, then to $t=1.0$ 
with $1024^2$ points and finally to $t=15.0$ with $512^2$ points. 
For this reason the energy spectra have different extents in Fourier
space at the different times $t = 0.0$ (dashed line), 0.1, 1.0, 
3.0, 6.0, \& 15.0. We include the full information, rather than just 
cutting off the plot at $k=100$, to demonstrate that our 
solutions are indeed fully resolved. The spectra show that energy 
is clearly transferred to $k<10$ in an inverse cascade, with the 
spectral peak shifting to $k=2$ but not maintaining the same 
amplitude. Some energy has also transferred to $k>20$ 
resulting in an overall spectrum comparable to a forward homogeneous 
cascade at late time. At $t=0.1$ the spectrum has a spectral index
of -5/2 for $k>10$. At late time this has steepened to $\sim -3$.
Between $t = 12$ and $t = 15$, the spectral index has stabilised.
The inverse cascade phenomenon becomes less pronounced at lower 
values of $R_B$ with no inverse cascade at all below $R_B = 300$.

\begin{figure}
\begin{center}
\includegraphics[angle=0,width=16cm]{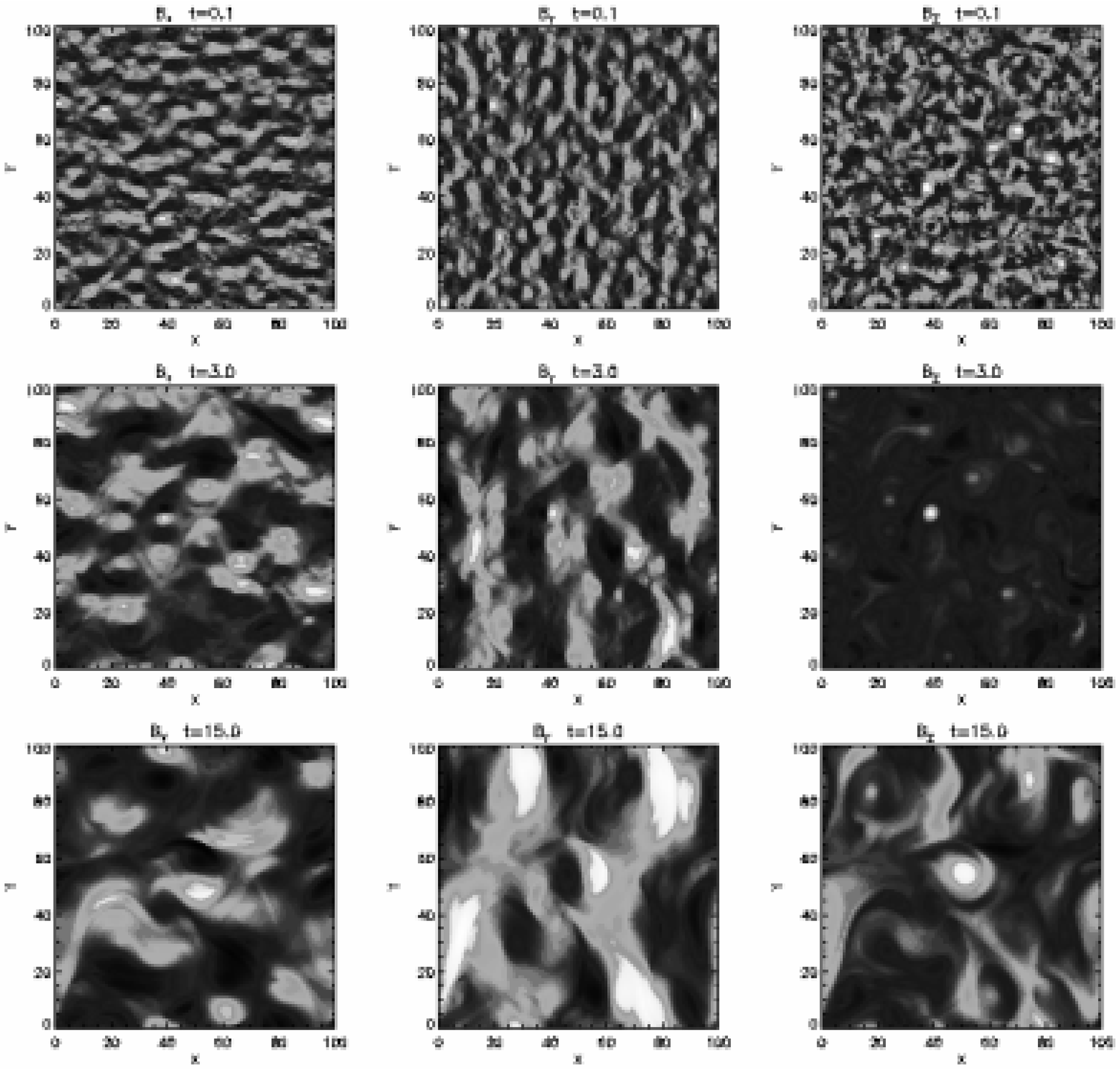}
\caption{Plots of the $R_B = 1000$ solution in real
configuration space at $t = 0.1$ (top row), $t=3.0$ 
(middle row) and $t = 15.0$ (bottom row). The fields
have been rescaled onto grids of $100 \times 100$ points.}
\label{invfield}
\end{center}
\end{figure}

In Figure \ref{invfield} we show the evolution of the field by 
including the real configuration space fields at times $t=0.1$, 3.0 and 15.0.
The initial field containing intermediate scale structure can be
seen to develop large scale structures, which unlike ordinary or 
MHD turbulence again appear to be relatively stationary.

Previous work has considered inverse cascade action in driven, rather
than decaying, 2D EMHD turbulence \cite{shaikh05}. There the authors
found a forward cascade of energy and an inverse cascade of mean squared
magnetic potential or anastrophy. Between the forcing lengthscale and the
artificial dissipative cutoff, the authors found a $k^{-7/3}$ energy spectrum
consistent with our results at $t=0.1$. It remains unclear though why
the spectrum steepens at late time. Previous work \cite{chertkov07} has found
a tendency toward energy condensation in forced 2D classical turbulence. There
the condensation is a finite size effect of the biperiodic box which occurs after
the standard inverse cascade reaches the size of the system. It leads to
the emergence of a coherent vortex dipole. It is important to note
that the dipole contains most of the injected energy and since we are
simulating decaying EMHD turbulence, we do not inject any energy which
could power the emergence of such a structure. The real field
in Figure \ref{invfield} shows no evidence of collimated dipole structure
and in fact a large number of isolated vortices can be seen, a characteristic
of fluid turbulence noted in \cite{shaikh05}. Here then we deduce we are 
seeing the first direct demonstration of the dual cascade phenomenon in 
decaying 2D EMHD turbulence. 

\section{Conclusions}

We have investigated the nature of decaying 2D EMHD turbulence with 
normal diffusivity and compared it with classical and MHD turbulence 
and studies of 2D EMHD with hyperdiffusivity. We have found
EMHD turbulence experiences an isotropic forward cascade of energy 
to higher wavenumber (smaller spatial scales) asymptotically
approaching $E_k \propto k^{-5/2}$ with increasing $R_B$ (inversely
proportional to a dissipation coefficient), in broad agreement with previous
studies. We have found there is no dissipative 
cutoff at the predicted wavenumber $k \sim R_B^{4/5}$ and argue this is
consistent with non-local coupling in Fourier space, the most important
result of this paper. Hyperdiffusivity 
has previously clouded this issue and introduced an artificial cutoff.
Only now, when we can avoid its use, has the true nature become clear.
We have also found that fully developed EMHD 
turbulence appears to be strongly structured, retaining a similarity
to the initial field at late time, very much unlike classical or MHD
turbulence and a point not noted in previous literature. 
Our study of EMHD turbulence with normal diffusivity has been found to display 
scale-dependent anisotropy in the presence of a uniform background 
field, in good agreement with previous studies employing hyperdiffusivity. Further,
our results support previous studies which found the strength of the anisotropy 
is linearly related to the external field strength. 
Finally, we have discovered that 
decaying EMHD turbulence is capable of yielding an inverse cascade, 
at least partially transferring magnetic energy from intermediate to 
large lengthscales. This result may 
be particularly significant for the magnetic fields of neutron stars, 
where the proto-neutron star that emerges from a supernova explosion 
may well have a primarily small-scale, disordered field. A Hall-induced
inverse cascade may then be a mechanism whereby it acquires a large-scale, 
ordered field.

\vskip0.5cm
\noindent

{\bf Acknowledgements}\
We thank Steve Tobias for his assistance in benchmarking the code
and an anonymous referee for their insightful comments enabling us 
to improve the paper. This work was supported by the Science \& 
Technology Facilities Council [grant number PP/E001092/1].

\label{lastpage}


\begin{thebibliography}{300}

\bibitem[\protect\citeauthoryear{Goldreich \& Reisenegger}{1992}]{goldreich92} P. Goldreich, A. Reisenegger, Astrophys. J. {\bf 395}, 250 (1992).
\bibitem[\protect\citeauthoryear{Hollerbach \& Rudiger}{2002}]{hollerbach02} R. Hollerbach, G. R\"{u}diger, Mon. Not. Roy. Astron. Soc. {\bf 337}, 216 (2002).
\bibitem[\protect\citeauthoryear{Biskamp, Schwarz \& Drake}{1996}]{biskamp96} D. Biskamp, E. Schwarz, J.F. Drake, Phys. Rev. Lett. {\bf 76}, 1264 (1996).
\bibitem[\protect\citeauthoryear{Biskamp et al.}{1999}]{biskamp99} D. Biskamp, E. Schwarz, A. Zeiler, A. Celani, J. Drake, Phys. Plasmas {\bf 6}, 751 (1999).
\bibitem[\protect\citeauthoryear{Dastgeer et al.}{2000}]{dastgeer00} S. Dastgeer, A. Das, P. Kaw, P. Diamond, Phys. Plasmas {\bf 7}, 571 (2000).
\bibitem[\protect\citeauthoryear{Dastgeer \& Zank}{2003}]{dastgeer03} S. Dastgeer, Zank G.P., Astrophys. J. {\bf 599}, 715 (2003).
\bibitem[\protect\citeauthoryear{Cho \& Lazarian}{2004}]{cho04} J. Cho, A. Lazarian, Astrophys. J. {\bf 615}, L41 (2004).
\bibitem[\protect\citeauthoryear{Shaikh \& Zank}{2005}]{shaikh05} Sheikh D., Zank G.P., Phys. Plasmas {\bf 12}, 122310 (2005)
\bibitem[\protect\citeauthoryear{Frigo \& Johnson}{2005}]{frigo05} M. Frigo, S.G. Johnson, Proc. of the IEEE {\bf 93}, 216 (2005).
\bibitem[\protect\citeauthoryear{Berger}{1997}]{berger97} M.A. Berger, J. Geophysical Research {\bf 102}, 2637 (1997).
\bibitem[\protect\citeauthoryear{Kolmogorov}{1941}]{kolmogorov41} A.N. Kolmogorov, Proc. USSR Acad. Sciences {\bf 30}, 299 (1941) (Russian). Proc. Roy. Soc. A {\bf 434}, 9 (1980) (English).
\bibitem[\protect\citeauthoryear{Kraichnan \& Montgomery}{1980}]{kraichnan80} R.H. Kraichnan, D. Montgomery, Rep. Prog. Phys. {\bf 43}, 547 (1980).
\bibitem[\protect\citeauthoryear{Shebalin et al.}{1983}]{shebalin83} J.V. Shebalin, W.H. Matthaeus, D. Montgomery, J. Plasma Phys. {\bf 29}, 525 (1983).
\bibitem[\protect\citeauthoryear{Oughton et al.}{1998}]{oughton98} S. Oughton, W.H. Matthaeus, S. Ghosh, Phys. Plasmas {\bf 5}, 4235 (1998).
\bibitem[\protect\citeauthoryear{Galtier}{2006}]{galtier06} S. Galtier, J. Plasma Phys. {\bf 72}, 721 (2006).
\bibitem[\protect\citeauthoryear{Kraichnan}{1967}]{kraichnan67} R.H. Kraichnan, Phys. Fluids {\bf 10}, 1417 (1967).
\bibitem[\protect\citeauthoryear{Chertkov et al.}{2007}]{chertkov07} M. Chertkov, C. Connaughton, I. Kolokolov, V. Lebedev, Phys. Rev. Lett. {\bf 99}, 084501 (2007).

\end{thebibliography}
\end{document}